\documentclass{article}

\usepackage{PRIMEarxiv}

\usepackage[utf8]{inputenc} 
\usepackage[T1]{fontenc}    
\usepackage{hyperref}       
\usepackage{url}            
\usepackage{booktabs}       
\usepackage{amsfonts}       
\usepackage{nicefrac}       
\usepackage{microtype}      
\usepackage{lipsum}
\usepackage{fancyhdr}       
\usepackage{graphicx}       
\usepackage{amssymb}      
\usepackage{enumerate}
\usepackage{booktabs}
 
\graphicspath{{media/}}     

\usepackage[usenames]{color}

\def\ln{\mathop{\rm ln}\nolimits}

\title{Interpreting deviations between AR-VTG and GR}

\author{Roberto Dale \footnote{} \\
Departament d'Estad\'{\i}sica, Matem\`atiques i Inform\`atica and Centre of Operations Research (CIO), \\
Universitat Miguel Hern\'andez \\
Elx, Alacant, Spain \\
\texttt{rdale@umh.es} \\
\And
Diego S\'aez\\
Departament d'Astronomia i Astrof\'isica \\ Universitat de Val\`encia \\
Burjassot, Val\`encia, Spain \\
Observatori Astron\`omic,  Universitat de Val\`encia \\ 
Paterna, Val\`encia, Spain \\
\texttt{diego.saez@uv.es} 
}

\begin{document}
\maketitle

\begin{abstract}
The Cosmic microwave background (CMB) anisotropies predicted by
two cosmological models are compared, one of
them is the standard model of general relativity with cold dark matter
and cosmological constant, whereas the second
model is based on a consistent vector-tensor theory of gravitation
explaining solar system and cosmological observations.
It is proved that the resulting differences 
-between the anisotropies of both models- are due
to the so-called late integrated Sachs Wolfe effect and,
consequently, cross correlations between
maps of CMB temperatures and tracers of the dark matter distribution
could be used in future to select one of the above models. The role
of reionization is analysed in detail.
\end{abstract}



\maketitle

\footnotetext{Corresponding author \texttt{rdale@umh.es}.}

%

\section{Introduction}
\label{sec:1}

In Ref.~\cite{planck2015_14} an analysis of modified gravity theories, which takes in account 
current CMB data, was presented. As it is stated in this paper: "Those models which are close to
$\Lambda {\rm{CDM}}$ are in broad agreement with current constraints on the background cosmology, 
but the perturbations may still evolve differently and hence it is important to test their predictions against CMB data". 
In agreement with these comments, present paper studies the case of another successfully modified gravity theory 
that is close to the $\Lambda {\rm{CDM}}$ model.

Any vector-tensor gravity (VTG) theory involves the metric tensor $g^{\mu \nu} $
and a vector field $A^{\mu} $. These fields are coupled to build up an appropriate
action leading to the basic equations via variational calculations.

There are many actions and VTG theories \cite{wil93,wil06} , 
here we focus our attention on one of these theories free from quantum ghosts and classical instabilities.
It has the same parameterized post-Newtonian limit as general relativity (GR) and
may explain cosmic microwave background (CMB) anisotropies as well as GR
\cite{dal09,dal12,dal14,dal17} . In this vector-tensor theory, the outer metric
corresponding to a spherically symmetric mass distribution has the same form as the well known 
Reissner-Nordstr\"{o}m solution of Einstein-Maxwell equations \cite{dal15} , whose source
is a charged 
spherically symmetric mass distribution. This implies that, in the VTG
theory (no charges),
there are repulsive gravitational forces at stellar scales, just as it occurs in 
Einstein-Maxwell theory for a charged star. These forces might affect 
neutron star structure and the gravitational collapse (to be studied elsewhere). 
Moreover, there is also
a gravitational cosmological repulsion as that due to the cosmological constant.
On account of these facts, this theory shall from now on be called 
attractive-repulsive vector-tensor gravity (AR-VTG).
As it was claimed in Ref.~\cite{dal17}, it is convenient to mention that this theory is not a particular case of 
the Generalized Proca Theories (GPT) \cite{hei14}  , which also involve a vector field $A^{\mu} $.

As it was discussed in Ref.~\cite{dal14}, the CMB anisotropies produced at $z>10$ are expected to 
be identical in GR and AR-VTG; nevertheless, close to $z=10$, some AR-VTG scalar cosmological modes
(see mode definitions in Refs.~\cite{bar80,huw97}), involved in the equations 
describing the evolution of the CMB photon distribution function (see Ref.~\cite{mb95}), begin to deviate 
from those of GR. These deviations might produce significant differences between the CMB anisotropies of GR 
and AR-VTG, but the redshift dependence of these differences requires numerical estimates which 
are performed below. Evidently, 
primary anisotropy is produced at $z>>10$ and; consequently, the differences between GR and AR-VTG
anisotropies at $z<10 $ must be due to some kind of secondary anisotropy. At these low redshifts various
effects are being produced: (i) the effect due to 
reionization (hereafter called R-effect), which is due to the interaction of
free electrons with CMB photons via Thompson scattering,
(ii) the so-called late integrated Sachs-Wolfe (LISW) effect due to large strictly linear scales, and
(iii) the Rees-Sciama effect produced by smaller nonlinear scales.

The Rees-Sciama effect is too small to be detected \cite{ful17} ; hence, we focus our attention 
on the R and the LISW effects \cite{hu94,zal95} . In GR, these effects may be estimated with the code 
CAMB \cite{lew00} , whereas a new code (hereafter VTCAMB) has been specially designed -by us- to carry out
the corresponding estimations in the context of AR-VTG (see Ref.~\cite{dal17}).

In this paper, $G$, $a$, $\tau$, and $z$ stands for the gravitation constant, the scale factor, the
conformal time, and the redshift, respectively. Our signature is (--,+,+,+). Greek 
indexes run from 0 to 3, while the latin ones from 1 to 3. Symbol $\nabla $ ($\partial $) stands
for a covariant (partial) derivative. Whatever the function $f$ may be, $\dot f$ stands 
for the partial derivative with respect to the conformal time.
The antisymmetric tensor $F_{\mu \nu} $ is defined by the relation 
$F_{\mu \nu} = \partial_{\mu} A_{\nu } - \partial_{\nu} A_{\mu }$. It has nothing to 
do with the electromagnetic field. 
Quantities $R_{\mu \nu}$, $R$, and $g$ are the covariant components of the Ricci 
tensor, the scalar curvature and
the determinant  of the matrix $g_{\mu \nu}$ formed by the covariant components 
of the metric, respectively. 
Units are chosen in such a way that the speed of light, $c$,
takes on the value $c= 1$. Quantity $\Delta_{\ell} = \ell (\ell +1) C_{\ell} /2\pi $, given in $\mu K^{2}$,
is considered as a measure of CMB angular power spectra. This quantity and other ones 
depending on it are represented in various 
Figures.
 
This paper is structured as follows: The AR-VTG theory is summarized in Sect. \ref{sec:2}.
The origin of the $\Delta_{\ell}$ deviations between GR and AR-VTG is analysed in Sect. \ref{sec:3},
and the variation of these deviations with the redshift is studied in Sect.\ref{sec:4}. Finally,
section \ref{sec:5} displays our main conclusions and an appropriate discussion.


\section{AR-VTG foundations.}   
\label{sec:2}

\subsection{Generalities.}   
\label{sec:21}
AR-VTG is particular parameterization of the general unconstrained VTG proposed by Will, Nordtvedt and Hellings \cite{wil72,hel73}
in early 1970s. All these theories were based on the action \cite{wil93}:
\begin{eqnarray}
I &=& \frac{1}{{16\pi G}}\int (R + \omega {A_\mu }{A^\mu }R + \eta {R_{\mu \nu }}{A^\mu }{A^\nu }  \nonumber \\ 
& &
- \varepsilon F_{\mu \nu}F^{\mu \nu} + \gamma {\nabla _\nu }{A_\mu }{\nabla ^\nu }{A^\mu } + {L_m})\sqrt { - g} \,{d^4}x ,
\label{action_vt}
\end{eqnarray}
where $\omega$, $\eta$, $\varepsilon$ and $\gamma$ are arbitrary parameters and $L_m$ 
is the matter Lagrangian density, which couples matter with the fields of the VTG theory.

There is a detailed analysis of the viability for VTG’s theories in Ref.~\cite{bel09} which determines the theories that may deserve our attention.
This studio includes the calculation of the propagation speeds for the different perturbation modes and also the conditions for the absence of quantum ghosts.
Regarding Ref.~\cite{bel09} the parameterization $\omega = 0$, $\eta = \gamma$, leaving $\varepsilon$ arbitrary is a viable set of theories 
in terms of classical stability, local gravity constraints and absence of ghosts. 
Moreover the perturbations of mentioned model propagate at the speed of light which leads to the absence of classical unstable modes.
AR-VTG theory is obtained when previous parameterization settings are applied to the general unconstrained VTG.

Let us now briefly summarize  
the AR-VTG basic equations, which were 
derived in Refs.~\cite{dal09,dal12} from an appropriated 
action, which is a particularization of the general vector-tensor action 
given in Ref.~\cite{wil93}. The resulting field equations are:

\begin{equation}  
G^{\mu \nu} = 8\pi G (T^{\mu \nu}_{GR} + T^{\mu \nu}_{VT}) ,
\label{fieles_vt}
\end{equation}   
\begin{equation}
2(2\varepsilon - \gamma)\nabla^{\nu} F_{\mu \nu} = J^{^{A}}_{\mu} ,
\label{1.3_vt}
\end{equation} 
where $G^{\mu \nu}$ is the 
Einstein tensor, $T^{\mu \nu}_{GR}$ is the GR energy momentum tensor,
$J^{^{A}}_{\mu} \equiv -2 \gamma \nabla_{\mu} (\nabla \cdot A)$ 
with $\nabla \cdot A = \nabla_{\mu} A^{\mu} $, and 
\begin{eqnarray}
T^{\mu \nu}_{VT} &=& 2(2\varepsilon - \gamma) [F^{\mu}_{\,\,\,\, \alpha}F^{\nu \alpha}
- \frac {1}{4} g^{\mu \nu} F_{\alpha \beta} F^{\alpha \beta}] \nonumber \\ 
& &
-2\gamma [ \{A^{\alpha}\nabla_{\alpha} (\nabla \cdot A) + \frac {1}{2}(\nabla \cdot A)^{2}\}
g^{\mu \nu} \nonumber \\
& &
-A^{\mu}\nabla^{\nu} (\nabla \cdot A) - A^{\nu}\nabla^{\mu} (\nabla \cdot A)] .
\label{emtee_vt}
\end{eqnarray}  
Equation (\ref{1.3_vt}) leads to the following conservation law 
\begin{equation}
\nabla^{\mu} J^{^{A}}_{\mu} = 0  
\label{confic}
\end{equation}
for the fictitious current $J^{^{A}}_{\mu}$. Moreover, the 
conservation laws $\nabla_{\mu} T^{\mu \nu}_{GR} = 0 $ and  
$\nabla_{\mu} T^{\mu \nu}_{VT} = 0 $ are satisfied by any solution 
of Eqs.~(\ref{fieles_vt}) and~(\ref{1.3_vt}).

The pair of parameters
($\varepsilon$, $\gamma$) must satisfy the inequality $2\varepsilon - \gamma > 0$ to prevent 
the existence of quantum ghosts and unstable modes in AR-VTG (see Ref.~\cite{bel09}). 

\subsection{The background cosmology.}   
\label{sec:22}

Let us now consider a flat uncharged homogeneous and isotropic background universe with matter and radiation
where the isentropic uncharged perfect fluid is characterized by an energy density $\rho_B$ and a pressure  $p_B$
(subscript B refers to background quantities). In this flat background using conformal time, the metric can be written in the form:
\begin{equation}  
d{s^2} = {a^2}( \tau)[ { - d{\tau ^2} + d{r^2} + {r^2}d{\theta ^2} + {r^2}{{\sin }^2}\theta d{\phi ^2}}] .
\label{ds2_bc}
\end{equation} 

Furthermore, the covariant components of the vector field are $(A_{0B}( \tau),0,0,0)$ and tensor $F_{\mu \nu}$ vanish. 
On the other hand it is worthwhile to notice that the relation ${\nabla _\mu }T_{VT}^{\mu \nu } = 0$ is satisfied (see Ref.~\cite{wil93}) 
so matter and radiation evolve as in the standard Friedmann-Robertson-Walker model of GR (immediately ${\nabla _\mu }T_{GR}^{\mu \nu } = 0$ is obtained).

Taking into account $F_{\mu \nu} = 0$, Eq.~(\ref{1.3_vt}) leads to
\begin{equation}
J^{^{A}}_{\mu} \equiv -2 \gamma \nabla_{\mu} (\nabla \cdot A) = 0 ,
\label{eqJA_cos}
\end{equation}
and then the quantity $( {\nabla  \cdot A} )_B$ is a constant and, consequently, tensor $T_{\mu \nu }^{VT}$ has the same form as the energy-momentum tensor
corresponding to vacuum; namely, one has $T_{\mu \nu }^{VT} =  - \rho _B^{VT}{g_{\mu \nu }}$, where 
$\rho _B^{VT} = {\textstyle{\gamma  \over 2}}(\nabla  \cdot A)_B^2 = constat \ne 0$ is the energy density due de vector field. 
This means that the resulting theory is equivalent to GR plus a cosmological constant. In terms of the component
${A_{0B}}( \tau)$ the equation (\ref{eqJA_cos}) is 
\begin{equation}
\ddot A_{0B} + 2 A_{0B} \bigg( \frac{\ddot a}{a} - 3 \frac{{\dot a}^2}{a^2} \bigg) = 0 ,
\label{eqA0B}
\end{equation}
while equations \ref{fieles_vt} are
\begin{equation}
3\frac{{{{\dot a}^2}}}{{{a^2}}} = 8\pi G{a^2}( {{\rho _B} + \rho _B^{VT}})
\label{eqroAB}
\end{equation}
and
\begin{equation}
- 2\frac{{\ddot a}}{a} + \frac{{{{\dot a}^2}}}{{{a^2}}} = 8\pi G{a^2}( {{p_B} + p_B^{VT}}) ,
\label{eqpAB}
\end{equation}
where
\begin{equation}
\rho _B^{VT} =  - p_B^{VT} = {\textstyle{1 \over 2}}\gamma ({\nabla \cdot A})_B^2 = \frac{\gamma}{{2{a^4}}}{({{{\dot A}_{0B}} + 2\frac{{\dot a}}{a}{A_{0B}}})^2} .
\label{eqropAB}
\end{equation}
Hence the equation of state is  $w^{VT} \equiv p^{VT}/\rho ^{VT} = -1$ , so due the energy density of the evolving field   is constant, we can state that this field acts as a cosmological constant.
Obviously the condition $\gamma >0$ must be required to have a positive $A^{\mu} $ energy density 
in the background universe (see Ref.~\cite{dal14}); hence, taking into account the previous inequality $2\varepsilon - \gamma > 0$, 
the inequalities $\varepsilon > \frac {\gamma}{2}>0 $ must be satisfied.

The necessary initial values for the integration are obtained at high redshift $( z_{\rm{in}} \sim {10}^8)$ during the radiation dominated era,
where it’s found that $a \propto \tau$ and ${A_{0B}}( \tau ) \propto {\tau ^3}$ satisfy the above background field equations. 
At $z = z_{\rm{in}}$ one finds:
\begin{equation}
\tau_{\rm{in}} = \bigg(\frac{\dot a}{a} \bigg)_{\rm{in}}^{ - 1}\quad ,( A_{0B})_{{\rm{in}}} = 
- \frac{{{{( {\nabla  \cdot A} )}_B}}}{{5( {1 + {z_{{\rm{in}}}}})}} \bigg( {\frac{{\dot a}}{a}} \bigg)_{{\rm{in}}}^{ - 1} .
\label{eqA0Bin}
\end{equation}
Subscript "in" of previous quantities refers to initial values at $z = z_{\rm{in}}$.

\subsection{The cosmological perturbations.}   
\label{sec:23}

In order to describe cosmological perturbations, the formalism summarized in Ref.~\cite{bar80} (see also Ref.~\cite{huw97}) is used.
In this formalism there are three types of perturbations whose evolution is independent during the linear regime: scalar, vector, and tensor fluctuations.
There are no tensor modes involved in the expansion of the introduced vector field, so the existing ones satisfy the same equations as in GR.
Formally the vector fluctuations are as in Einstein-Maxwell. The main reason lies in the fact that the action (\ref{action_vt}) is full equivalent to 
\begin{equation}
I = \frac{1}{16\pi G}\int {[ {R + ( {{\textstyle{1 \over 2}}\gamma  - \varepsilon } )F_{\mu \nu }F^{\mu \nu } + \gamma ( {\nabla  \cdot A})^2 + {L_m}}]\sqrt { - g} \,{d^4}x} \,
\label{actionbis_vt}
\end{equation}
for the parameterization $\omega =0$, $\eta = \gamma$, and this action differs in essence, from the Einstein-Maxwell one, in the term proportional to  $({\nabla  \cdot A})^2$,
which is an scalar. 

At linear level the vector perturbation of the vector field can be written $A_\mu ^{(1)} = \big(0,{A^{(1)}}Q_i^{(1)} \big)$, where $Q_i^{(1)}$ 
are the vector harmonics which are a solution of the Helmholtz equation ${\nabla ^2}Q_i^{(1)} = - {k^2}Q_i^{(1)}$
(see Ref.~\cite{bar80}), and $k$ is the wave number that sets the spatial scale of the perturbation. The evolution equation for the vector modes amplitude $A^{(1)}$ is \cite{dal17}
\begin{equation}
\ddot A^{(1)} + k^2 A^{(1)} = 0 .
\label{eqA1}
\end{equation}
It's interesting realize that this mode is uncoupled from the rest of mode equations, so it has no impact on the evolution of the other modes. 

Finally, all the scalar modes of GR are also involved in AR-VTG but, in order to describe the scalar modes associated to the field $A_\mu$,
we introduce a new gauge invariant scalar mode \cite{dal12,dal14} $({\nabla \cdot A})^{(0)}$ which is the first order term in the harmonic 
expansion of the scalar function $({\nabla \cdot A})$ defined as follows:
\begin{equation}
(\nabla  \cdot A) = (\nabla  \cdot A)_B + (\nabla \cdot A)^{(0)}Q^{(0)} ,
\label{eqperdivA}
\end{equation}
where the scalar harmonic $Q^{(0)}$ is a solution of the Helmholtz equation  ${\nabla ^2}Q^{(0)} = - {k^2}Q^{(0)}$
(see Ref.~\cite{bar80}). There are no more independent AR-VTG scalar modes. From Eq.~(\ref{eqJA_cos}),
following uncoupled differential equation for the new mode is obtained:
\begin{equation}
{(\nabla \cdot A)^{(0)}}^{\bullet  \bullet} + 2aH  {(\nabla \cdot A)^{(0)}}^{\bullet} + k^2 (\nabla \cdot A)^{(0)} = 0 .
\label{eqdivA0}
\end{equation}
This equation just involves, apart from the new AR-VTG scalar mode and its derivatives, the background functions $a(\tau)$,
$H(\tau) \equiv {\dot a} / a^2$ and the wave number. It's convenient to write scalar perturbation equations in the synchronous gauge 
and in terms of the scalar modes defined in Ref.~\cite{mb95}, the reason is because those are the functions and gauge used by the original CAMB 
code and the modified one VTCAMB. There are just AR-VTG corrections terms to the standard GR in equations (21a)--(21c) derived in Ref.~\cite{mb95}, 
this set of modified equations are:
\begin{equation}
{k^2}\tilde \eta  - {\textstyle{1 \over 2}}aH{\kern 1pt} \dot {\tilde h} = -4\pi G \big [ {a^2}{\rho _B}\tilde \delta + \Psi_A^{(0)} \big] ,
\label{eqperesc01}
\end{equation}
\begin{equation}
{k^2}\dot {\tilde \eta} = 4\pi G \bigg [{a^2}(\rho_B + p_B) \tilde \theta + \frac{\gamma }{8\pi G}{k^2}A_{0B} ({\nabla \cdot A})^{(0)} \bigg] ,
\label{eqperesc02}
\end{equation}
\begin{equation}
\ddot {\tilde h} + 2aH \dot {\tilde h} -2 {k^2}\tilde \eta = -24\pi G \big[{a^2}p_B \pi_L - \Psi_A^{(0)}\big],
\label{eqperesc03}
\end{equation}
where $\Psi_A^{(0)} \equiv - \frac{\gamma }{8\pi G} \big [{a^2}{({\nabla \cdot A})}_B ({\nabla  \cdot A})^{(0)} + A_{0B} {({\nabla \cdot A})}^{(0) \bullet } \big]$. 
In the above equations $\tilde \eta$ and $\tilde h$ are the scalar modes related with the metric while $\tilde \delta$, $\tilde \theta$ and $\pi_L$ are the scalar modes related with fluid (density fluctuation, divergence of fluid velocity and the isotropic pressure perturbation, respectively). The same functions (without the tilde) and its definitions are found in Ref.~\cite{mb95}, 
while the $\pi_L$ one can be located in Ref.~\cite{bar80}. The rest of scalar modes equations remain unaltered (see Eqs. (92) and subsequent in paper Ref.~\cite{mb95}).

As in previous subsection, initial conditions equations for the AR-VTG scalar modes at redshift $( z_{\rm{in}} \sim {10}^8)$
in the radiation dominated era, are obtained. In GR the initial conditions equations involve a normalization
constant named $C$ at Ref.~\cite{mb95} while in AR-VTG it involves an extra one we call $D$, that is related 
with the initial spectrum of the new scalar mode in the following manner: $(\nabla  \cdot A)^{(0)} = D k^4$. 
The complete set of modified initial conditions equations can be found in  Ref.~\cite{dal17} labelled as (2.23).


\section{CMB anisotropy differences between GR and AR-VTG.}   
\label{sec:3}

\subsection{Numerical computational fitting.}   
\label{sec:31}

In order to fit observational data with the model predictions an adapted and modified version of the standard well known codes COSMOMC \cite{lew02} (hereafter VTCOSMOMC) 
and CAMB have been used. There are a lot of tasks related with the adaptation and modifications of the codes as the inclusion of the background and 
scalar modes extra field equations, and its corresponding initial conditions equations, the modification of the scalar modes equations including the new terms and 
again its modified corresponding initial conditions equations. But this is not enough, the introduction of the new parameter implies changes at original COSMOMC (the Markov sample chains generator software) code, modifications related with the integration steps in wave number $k$ and in time, numerical issues and other technical concerns.

As related in the above subsection there is a new constant $D$  whose absolute value $|D|$ normalizes the spectrum of the vector field $A_\mu$ (its divergence) scalar cosmological perturbations. This is the new parameter that has been added in the different fitting calculations. Apart of the mentioned above, we include six GR parameters to conform a minimal base model to fit, those parameters are: $\Omega_b h^2$, $\Omega_{DM} h^2$, $\tau_{re}$, $n_s$, $\log [10^{10} A_s]$, and $\theta$, where  $\Omega_b$ and $\Omega_{DM}$
are the density parameters of baryons and dark matter, respectively, $h$ is the reduced Hubble constant, $\tau_{re}$ is the reionization optical depth, 
$n_s$ is the spectral index of the power spectrum of scalar modes, and $A_s$ is the normalization constant of the same spectrum whose form is $P(k) = A_s k^{n_s}$, 
finally, the parameter $\theta$ (angular acoustic scale) is defined by the relation $\theta \times 10^{-2}= r_{s}(z_{*})/d_{A}(z_{*})$, where 
$r_{s}(z_{*})$ is the sound horizon at decoupling redshift $z_*$, and $d_{A}(z_{*}) $ is the angular diameter distance at the same redshift.
This minimal base model is expanded for some cases when tensor modes are included, in such a case, following additional parameters are considered:
$r_{0.05}$ (the primordial tensor to scalar initial amplitude at the pivot scale of $k_{0} = 0.05 Mpc^{-1}$), and $dn_s/d \ln k$ (running index).

\begin{table}[htbp]
\caption{Minimal fit for 6 (GR) + 1 ($D$) parameters in the AR-VTG model. The evidence sources used are Planck CMB anisotropies and WMAP polarization anisotropy at low $\ell \lesssim 23$.}
\centering
{\begin{tabular}{|c|c|c|c|} \hline\hline
Parameter & Best Fit & 68$\%$ Lower Limit &  68$\%$ Upper Limit \\ \hline

$D \times 10^{-8}$ & 1.596 & -2.149 & 2.149    \\ 
$\Omega_{b}h^{2}$ & 0.02216 & 0.02179 & 0.02235    \\
$\Omega_{DM}h^{2}$ & 0.1187 & 0.1169 & 0.1222    \\
$\tau_{re}$ & 0.0893 & 0.0749 & 0.1013    \\
$n_{s}$ & 0.9657 & 0.9535 & 0.9684   \\
$\log[10^{10}A_{s}]$ & 3.085 & 3.060 & 3.110   \\
$\theta$ & 1.0411 & 1.0407  & 1.0419   \\ 
\hline\hline
\end{tabular} \label{tab:1}}

\end{table}

\begin{figure*}[tb]
\begin{center}
\resizebox{1.\textwidth}{!}{%
\includegraphics{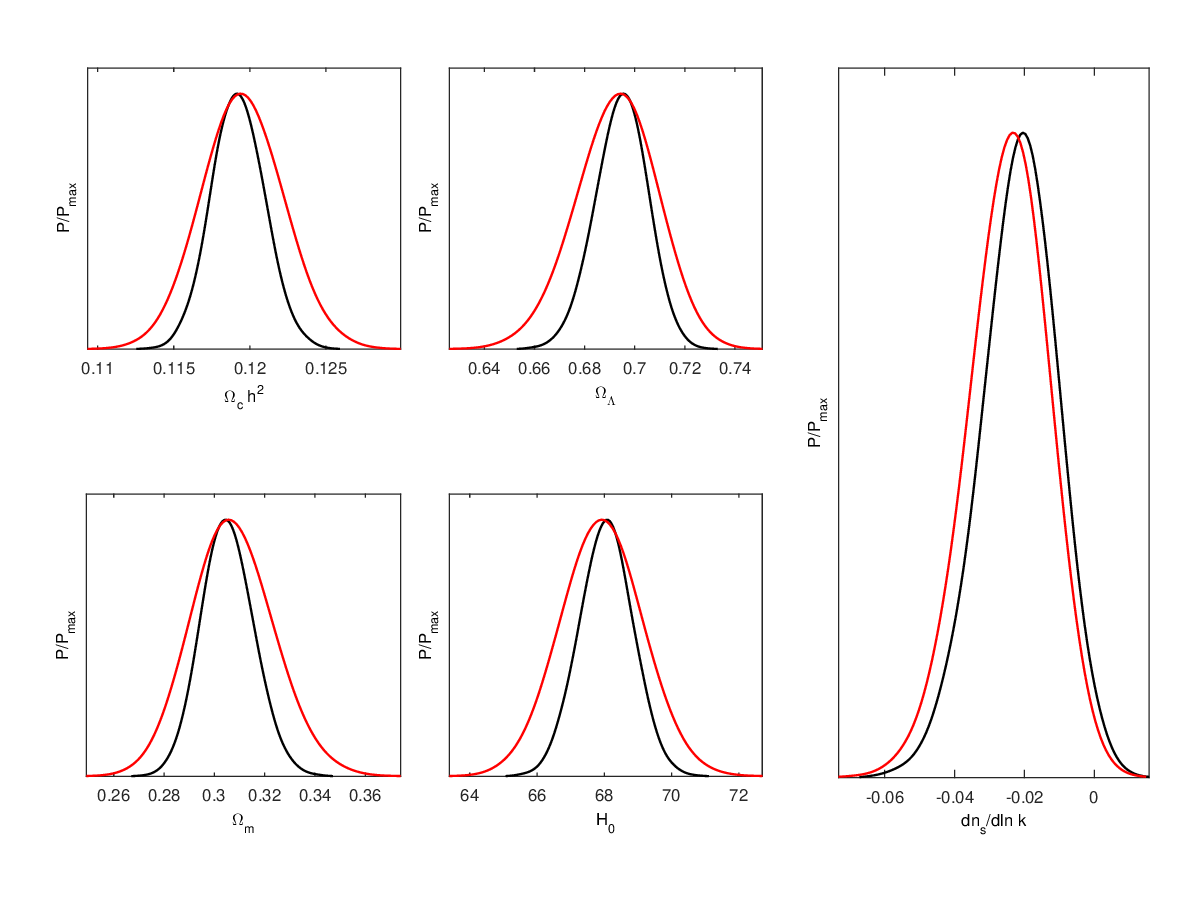}
}
\caption{Five panels representing the normalized to unity $(P/P_{\max})$ marginalized likelihood function for different standard GR parameters; 
$\Omega_m$ is the density parameter of matter and $H_0$ is the Hubble constant parameter. Black colour represents the function for GR while the red one is used for AR-VTG.}
\label{figuex01} 
\end{center}      
\end{figure*}

A variety combination of evidence sources that includes Ia supernovae (SNIa), WMAP 7 years CMB anisotropies (WP7), Planck CMB anisotropies (Planck), WMAP polarization anisotropy at low
$\ell \lesssim 23$ (WP), baryon acoustic oscillations (BAO), have been used.
A complete analysis of the results can be found at Ref.~\cite{dal14} and Ref.~\cite{dal17}, however let us summarize useful obtained information for the current studio. 
When the constant $D$ is considered as an additional parameter to be adjusted and a certain confidence level is assumed, we have found that, 
in AR-VTG fitting calculations, most GR parameters belong to intervals wider than those of the GR fitting calculations and, consistently, 
quantity $|D|$ takes on non-vanishing values. As a sample see  Fig.~\ref{figuex01} where the left four panels represent the normalized likelihood for different standard GR parameters 
(the best fit values, including the 68$\%$ confidence intervals can be found at Table \ref{tab:1}); at the stretched right panel 
the normalized likelihood function for the running index parameter, when tensor modes are included, is presented.
This is a pattern that is repeated when using different evidence data sources combinations and strongly suggests that a parameter $D \ne 0$ facilitates the 
adjustments between predictions and cosmological observations. 

\begin{figure*}[tb]
\begin{center}
\resizebox{1.\textwidth}{!}{%
\includegraphics{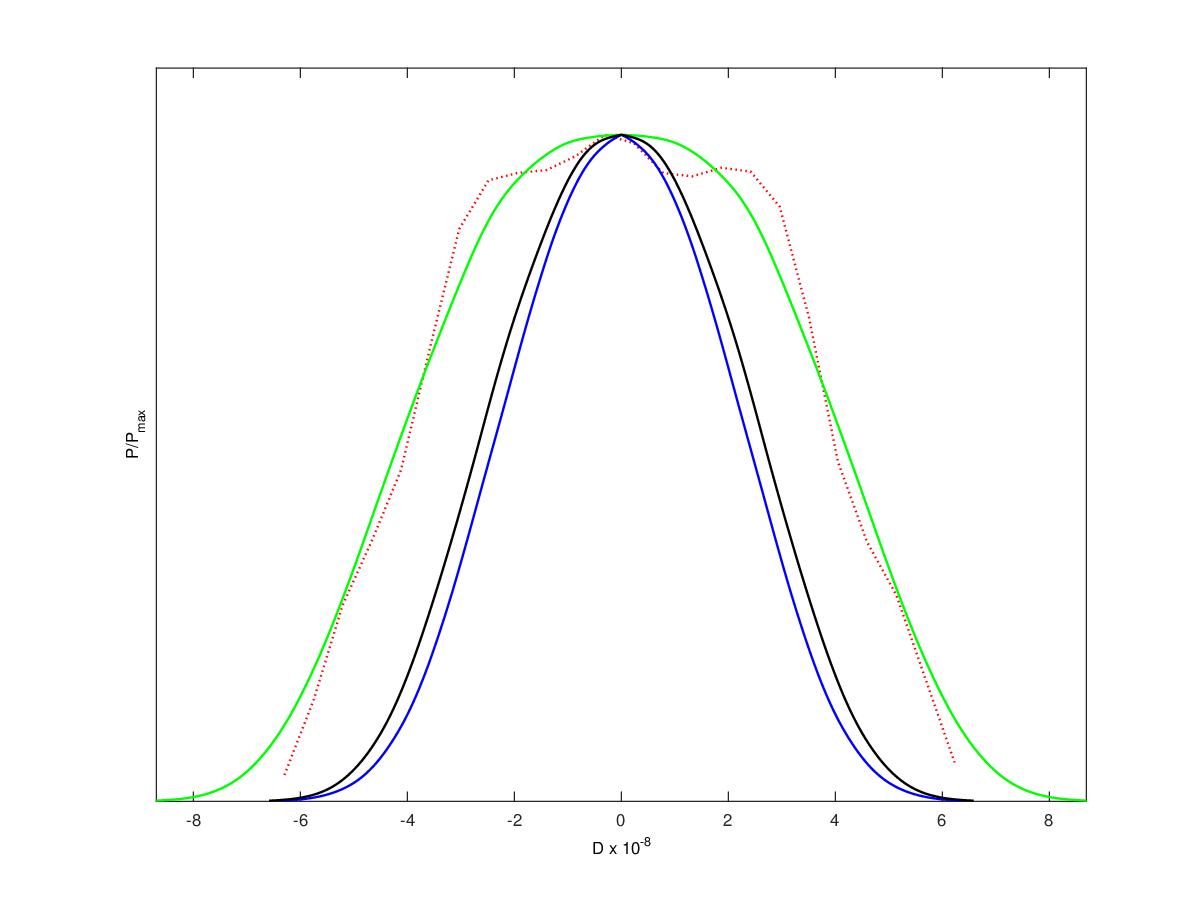}
}
\caption{Marginalized distribution functions normalized to unity for the parameter $D \times 10^{ - 8}$ in various fits. 
The dotted red curve was built with WMAP and SNIa data whereas blue solid, black, and green lines provide 
$(P/P_{\max})$ use Planck + WP just with scalar modes, Planck + WP including tensor modes without running index, 
and Planck + WP including tensor modes and running index, respectively.}
\label{figuex02} 
\end{center}      
\end{figure*}

Parameter $|D|$ plays a positive statistical role in the study of AR-VTG scalar perturbations. Depending on the set of evidence sources considered 
and the inclusion or not of tensor fluctuations (and the parameters related with it), different intervals for the  confidence level of $1 \sigma$ (68$\%$),  $2 \sigma$ (95$\%$), and
 $3 \sigma$ (99.7$\%$) are achieved. The Fig.~\ref{figuex02} represents a summary of mentioned results, for instance at $2 \sigma$ confidence level those are:
$[- 3.876,\;3.876]$ for just scalar perturbations and Planck + WP, $[- 4.005,\;4.005]$ same as previous one but including BAO, and $[- 5.442,\;5.442]$
when including tensor modes with running index and using Planck + WP as evidence sources.

\subsection{On the CMB anisotropy differences.}   
\label{sec:32}

Let us now describe the method used to analyse the origin of the 
CMB anisotropy differences between GR and AR-VTG.

Codes CAMB and VTCAMB are used
to estimate the R and LISW effects at $z \leq 10$.
Reionization is modelled by using the standard optical depth $\tau_{re} $ parameter. 

In any case, results from CAMB based on a minimal six parameters fit model are compared with the 
results obtained with VTCAMB for the same values of the six parameters plus a new parameter characteristic 
of AR-VTG. This parameter labelled $D$ will take on the value $4\times 10^{8}$,
which was proved to be compatible with CMB observations at the $2\sigma $ confidence level (see previous subsection and Ref.~\cite{dal17}).

We use two sets of six parameters obtained in previous papers -in the context of GR- 
to fit theoretical predictions and observations.
These sets are hereafter called $\Lambda {\rm{CDM{-}2013}}$ and $\Lambda {\rm{CDM{-}2015}}$. 
The six parameter values for $\Lambda {\rm{CDM{-}2013}}$ ($\Lambda {\rm{CDM{-}2015}}$)
are given in the first (second) data row of Table \ref{tab:2}. 

\begin{table}[ph]
\caption{Two minimal fits (six parameters) in the standard $\Lambda$CDM model.}
\centering
{\begin{tabular}{|c|c|c|c|c|c|c|} \hline\hline
Parameters & $\Omega_{b}h^{2}$ &  $\Omega_{DM}h^{2}$ & $\tau_{re}$ & $n_{s}$ & $\log[10^{10}A_{s}]$ & $\theta$ \\ \hline
$\Lambda {\rm{CDM-2013}}$ & 0.02209 & 0.1195 & 0.0927  & 0.9633  & 3.093 & 1.0415    \\ 
$\Lambda {\rm{CDM-2015}}$ & 0.02227 & 0.1184 & 0.067  & 0.9681  & 3.064 & 1.04106    \\ 
 \hline\hline
\end{tabular} \label{tab:2}}
\end{table}

Five of the six parameters given by CAMB -in the GR context- take on similar values 
whatever the observational data may be (WMAP, PLANCK, and so on); however, 
parameter $\tau_{re}$ depends on the CMB polarization data used 
in fit calculations. 
The $\Lambda {\rm{CDM{-}2013}}$ parameter values --obtained in Ref.~\cite{dal17}--
were derived by using Planck data about the CMB temperature 
distribution, plus WMAP data on CMB polarization anisotropy at low $\ell \lesssim 23$,
and the resulting  $\tau_{re}$ is close to 0.093 (first data row of Table \ref{tab:2});
nevertheless, the $\Lambda {\rm{CDM{-}2015}}$ parameter values, given in the second data row of Table \ref{tab:2},
were derived in Ref.~\cite{planck2015_13} by using both temperature and polarization Planck
data (see column 3 of Table 4 in this last reference); in this second case, the parameter value
$\tau_{re}$ is close to 0.067; namely, this parameter is rather smaller than that 
of the $\Lambda {\rm{CDM{-}2013}}$ fit, which is a consequence of important
differences in the polarization data. Both fits have been performed by using 
CAMB and COSMOMC codes. December 2013 (November 2015) versions were used 
to get the $\Lambda {\rm{CDM{-}2013}}$ ($\Lambda {\rm{CDM{-}2015}}$) parameter values.

As it has been stated in Sect. \ref{sec:1}, the differences between the 
CMB anisotropies of GR and AR-VTG must be due to a combination of the R and LISW effects
at $z \leq 10$. Reionization contributes to the LISW effect \cite{hu94} ($R1$-effect) and, moreover, it also creates 
anisotropy due to path mixing, Doppler effects due to motions in the reionization electron 
distribution, and so on ($R2$-effect). Our main goal is to disentangle the $R1$, $R2$, and LISW 
contributions to find out the nature of the aforementioned 
differences. Since there are well defined terms giving the LISW effect -in CAMB 
and VTCAMB- and these terms include reionization contributions ($R1$-effect), we can proceed as follows:

\begin{enumerate}[i]

\item Once a data row of Table \ref{tab:2} has been selected and the value $D=4\times 10^{8}$ has been fixed,
Codes CAMB and VTCAMB may be used to calculate the total $\Delta_{\ell}$ numbers in GR and AR-VTG,
respectively. 

\item For the same parameters, the LISW contribution to the $\Delta_{\ell}$ quantities may be easily 
calculated. This computation may be performed by integrating, from redshift $10$ to $0$,        
only the terms giving the LISW effect (including $R1$); namely, by canceling  
any other contribution to the CMB angular 
power spectrum, including $R2$ reionization effects (not contained in LISW); in this way, 
only the reionization contribution, $R1$, to the LISW effect is taken into account.

\end{enumerate}

\begin{figure*}[tb]
\begin{center}
\resizebox{1.\textwidth}{!}{%
\includegraphics{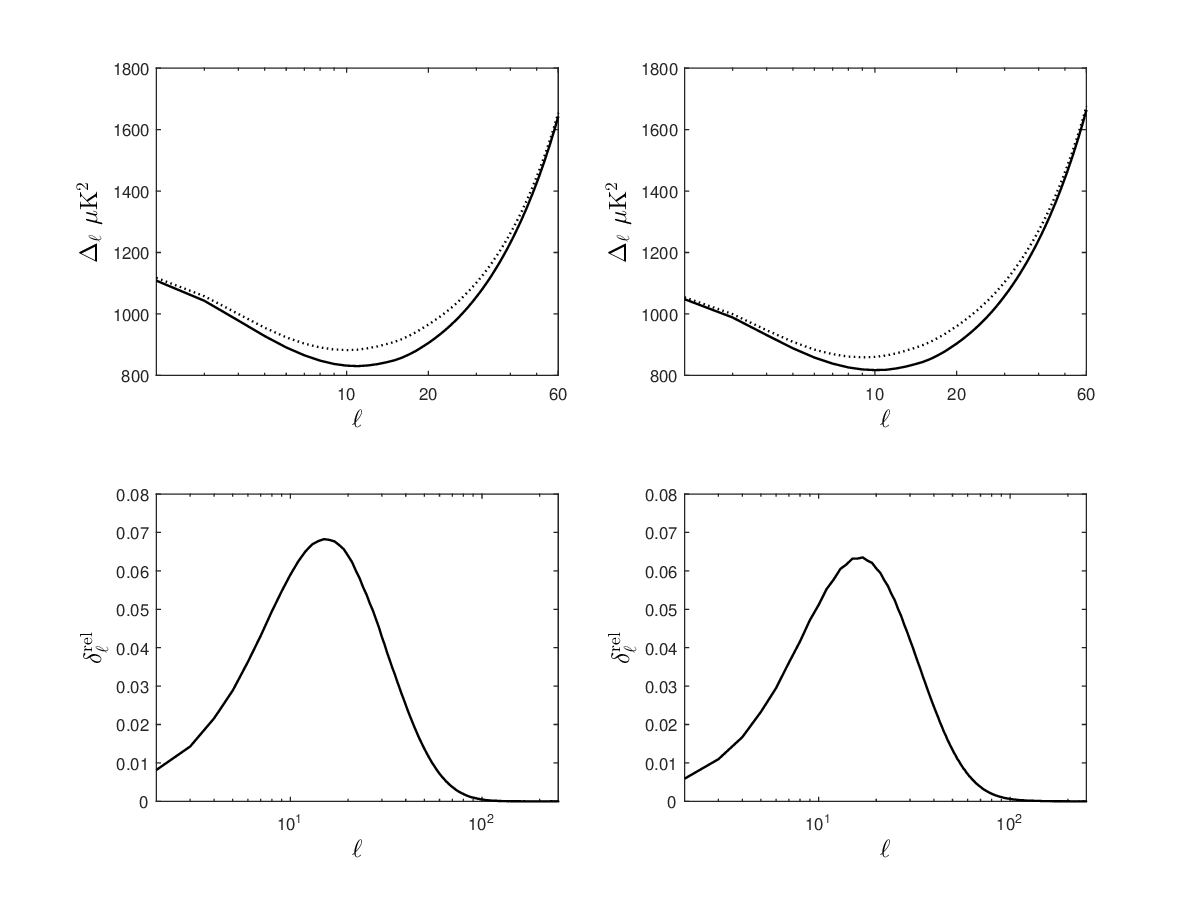}
}
\caption{Top left (right) panel shows the
total CMB angular power spectrum $\Delta_{\ell} $ for $\Lambda {\rm{CDM-2013}}$ 
($\Lambda {\rm{CDM-2015}}$) parameters. In these panels, solid (dotted) line corresponds to
the $\Lambda$CDM model of GR (AR-VTG with $D=4 \times 10^{8} $).
Bottom panels present the relative deviations, $\delta^{rel}_{\ell} $, between GR and 
AR-VTG, for the pairs of curves displayed in the top panels located 
at the same column.}
\label{figu0} 
\end{center}      
\end{figure*}

\begin{figure*}[tb]
\begin{center}
\resizebox{1.\textwidth}{!}{%
\includegraphics{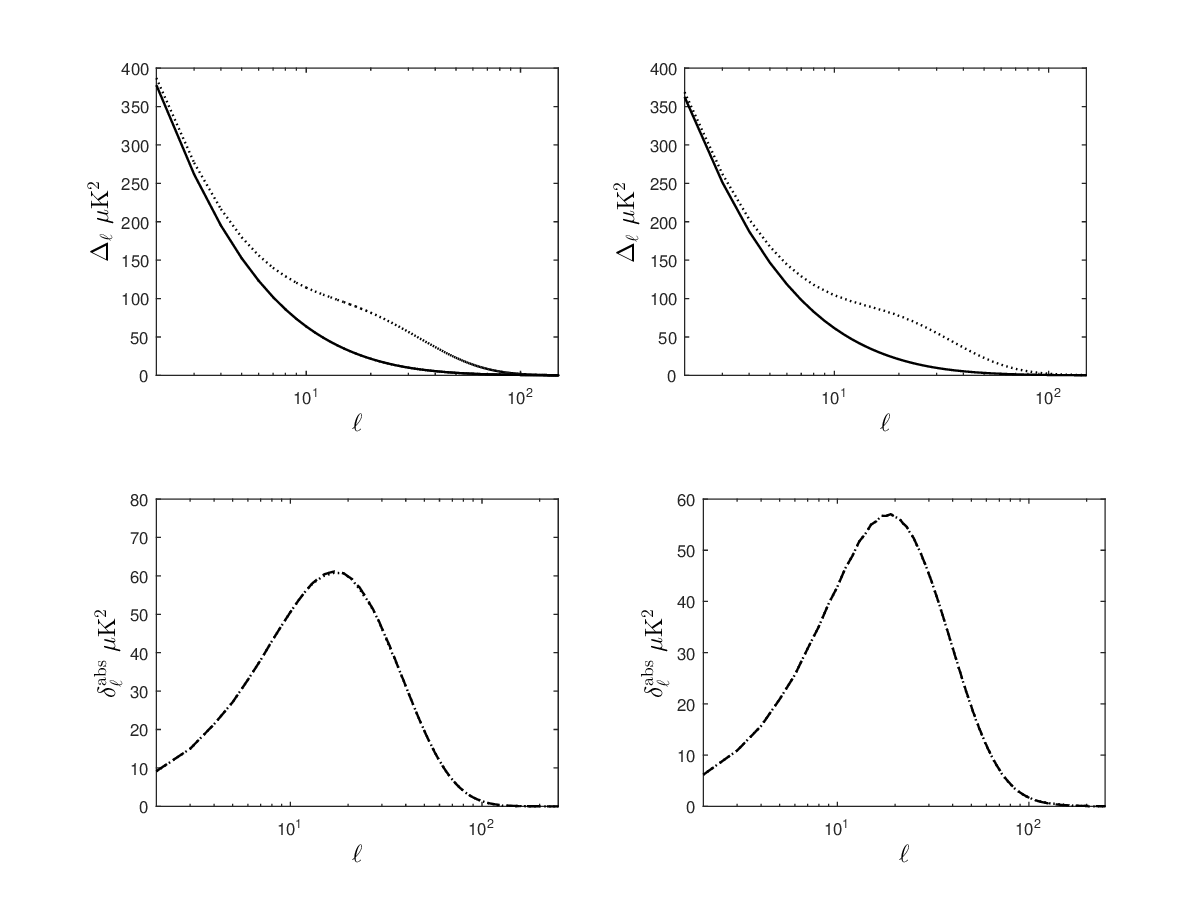}
}
\caption{Top panels are as those of Fig.~\ref{figu0}, but $\Delta_{\ell} $
is here the LISW contribution to the
total CMB angular power spectra. In bottom panels, the dashed (dotted) line
displays the absolute deviations, $\delta^{abs}_{\ell} $, between GR and 
AR-VTG, for the pairs of curves shown inside the  
top panel located in the same column of this Figure (of Fig.~\ref{figu0}).
}
\label{figu1} 
\end{center}      
\end{figure*}

\begin{figure*}[tb]
\begin{center}
\resizebox{1.\textwidth}{!}{%
\includegraphics{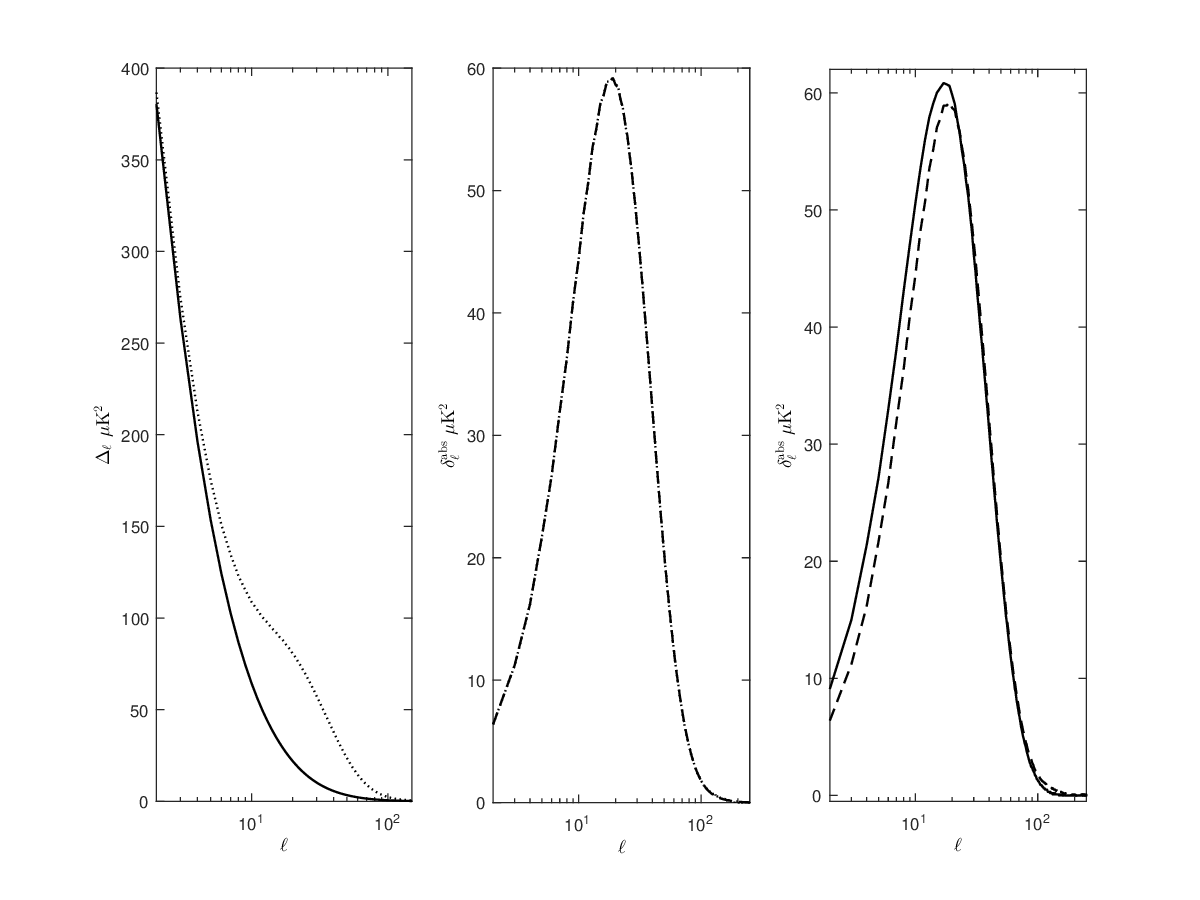}
}
\caption{Left and central panels have the same structure as the left top
and left bottom panels of Fig.~\ref{figu1}. In the panels of this Figure, 
quantities $\Delta_{\ell}$ and $\delta^{abs}$ 
have been calculated in the absence of reionization. In the right panel, solid (dashed) line 
gives LISW absolute deviations, $\delta^{abs}$, between GR and AR-VTG with (without)
reionization. All the curves correspond to the case $\Lambda {\rm{CDM{-}2013}}$.
}
\label{figu2} 
\end{center}      
\end{figure*}

\begin{figure*}[tb]
\begin{center}
\resizebox{1.\textwidth}{!}{%
\includegraphics{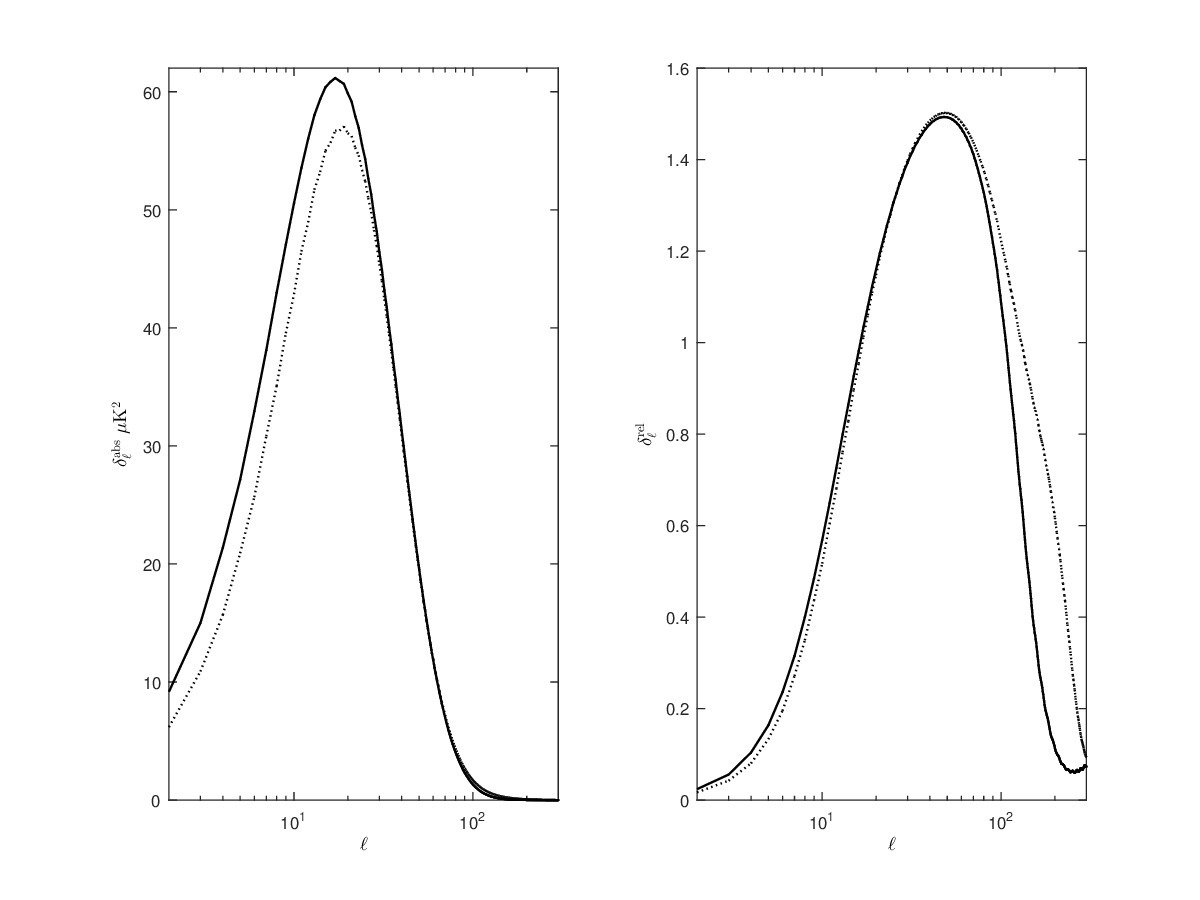}
}
\caption{Left (right) panel shows absolute (relative)
LISW deviations between GR and AR-VTG. Solid (dotted) lines  
correspond to $\Lambda {\rm{CDM{-}2013}}$ ($\Lambda {\rm{CDM{-}2015}}$).
}
\label{figu3} 
\end{center}      
\end{figure*}

Results obtained with this procedure may be used to calculate,
for the chosen parameters, both absolute and relative deviations between 
the GR and AR-VTG angular power spectra. These deviations may be 
estimated for the total $\Delta_{\ell}$ quantities, and also 
for the LISW contribution to the angular power spectrum. The absolute deviations are 
$\delta^{abs}_{\ell} = |\Delta_{\ell}(GR) - \Delta_{\ell}(AR-VTG)|$, whereas the relative 
ones are $\delta^{rel}_{\ell} =2 [|\Delta_{\ell}(GR) - \Delta_{\ell}(AR-VTG)|]/
[\Delta_{\ell}(GR) + \Delta_{\ell}(AR-VTG)]$. These deviations are presented in 
Figs.~\ref{figu0} to~\ref{figu3}.

The top panels of Fig.~\ref{figu0} exhibit the total $\Delta_{\ell} $ quantities 
obtained from the parameters of the 
first ($\Lambda {\rm{CDM{-}2013}}$, left) and second ($\Lambda {\rm{CDM{-}2015}}$, right) 
rows of Table~\ref{tab:2}. In these panels, 
solid (dotted) lines correspond to GR (AR-VTG theory with 
$D=4\times 10^{8}$). From the comparison of solid and dashed lines it follows that,
for $D=4\times 10^{8}$,  
there are deviations from GR in the $\ell $ interval represented. For $\ell > 60$,
both theories lead to almost the same angular power spectrum. The relative deviations 
are given in the bottom panels, where it can be seen that the values of these deviations 
are close to $0.06$ (6\%) 
for $\ell $ values in the interval (10,20),
being greater than $0.01$ (1\%) between $\ell=2$ and $\ell=60$. We will not
discuss the importance of these deviations, as it was already done 
in previous papers \cite{dal14,dal17} from the statistical point of view. 
We are only interested in the nature of these deviations, which are not 
either negligible or too large for the chosen $D$ value. The key question now is:
what kind of effect 
produces the relative deviations represented in Fig.~\ref{figu0}? To answer this question
we use Figs.~\ref{figu1} and~\ref{figu2}.
We have to emphasize that these two figures also reveals us that, while the aforementioned relative deviations in the $\ell$ 
interval are around a 6$\%$, when the isolate contribution due LISW is consider, those deviations reach, and even exceed, a 100$\%$.

If the absolute deviations corresponding to 
the total $\Delta_{\ell} $ quantities and those of the LISW contribution (including $R1$)
may be considered the same; namely, if the differences among these two 
absolute deviations are smaller than the numerical errors in the $C_{\ell} $ coefficients
due to CAMB and VTCAMB, it can be stated that the total deviations between 
GR and AR-VTG are fully due to the LISW effect; however, 
if these differences are greater than the expected 
numerical errors, a part of the total deviations between 
GR and AR-VTG would be due to reionization through effects which are not 
included in the total LISW ($R1$). 

Each of the top panels of Fig.~\ref{figu1} 
has the same structure as the corresponding top panel of Fig.~\ref{figu0};
only the displayed quantities are different in these figures:  
the total $\Delta_{\ell}$ coefficients in Fig.~\ref{figu0} and the contribution due to the LISW 
in Fig.~\ref{figu1}. In the bottom panels of this last Figure, there are two 
lines, the dashed line gives the absolute differences between GR and AR-VTG corresponding to the 
total $C_{\ell} $ coefficients represented in the top panels of Fig.~\ref{figu0},
whereas the dotted line displays the absolute differences of the LISW contributions given in the top panels
of Fig.~\ref{figu1}. The coincidence of these lines, which are indistinguishable to the eye, suggests us that 
the total deviations between GR and AR-VTG are essentially due to the LISW effect (see previous paragraph). See also 
section \ref{sec:5} for a detailed measurement of the relative deviations between the dotted and dashed lines 
of the bottom panels.

Let us now repeat our estimations of the total $\Delta_{\ell} $ quantities and the LISW 
contributions to them in the absence of reionization. For the sake of briefness,  
only the results corresponding to $\Lambda {\rm{CDM{-}2013}}$ 
parameters (first data row of Table \ref{tab:2}) are presented. 
Similar results have been verified for $\Lambda {\rm{CDM{-}2015}}$ parameters (second data row of Table \ref{tab:2}).
Since only the LISW effect is producing CMB anisotropy at $z < 10$, the  
total and LISW absolute deviations between GR and AR-VTG must
coincide. This fact has been verified and results are presented in Fig.~\ref{figu2}. 
Left and central panels of this figure have the same structure as top left and 
bottom left panels of Fig.~\ref{figu1}. The two curves of the central panel 
are almost identical (see section \ref{sec:5} for measurements of deviations), 
which confirms that the total anisotropy is a LISW effect.
Right panel shows LISW absolute deviations between GR and AR-VTG
with and without reionization. It is evident that reionization affects 
the LISW absolute deviations ($R1$-effect), but the absolute deviations of the total 
$C_{\ell} $ quantities are affected in such a way that
the two curves of the bottom left panel of Fig.~\ref{figu1} (with reionization)
are very similar, and those of the central panel of Fig.~\ref{figu2} (without reionization) 
are also almost identical.

Fig.~\ref{figu3} shows absolute (left) and relative (right) differences between GR and AR-VTG
for the LISW contributions to $\Delta_{\ell} $. Solid lines ($\Lambda {\rm{CDM{-}2013}}$ parameters)
and dotted lines ($\Lambda {\rm{CDM{-}2015}}$) do not coincide. 
A remarkable difference appears in both cases since reionization is 
very different for $\Lambda {\rm{CDM{-}2013}}$ and $\Lambda {\rm{CDM{-}2015}}$ parameters
(see the values of parameter $\tau_{re}$ in Table \ref{tab:2}).
In spite of this fact, the total $\Delta_{\ell} $ quantities are also different for $\Lambda {\rm{CDM{-}2013}}$
and $\Lambda {\rm{CDM{-}2015}}$ and, as it may be appreciate in the two 
bottom panels of Fig.~\ref{figu1}, where the dotted and dashed curves almost coincide 
for $\Lambda {\rm{CDM{-}2013}}$ (left) and also for $\Lambda {\rm{CDM{-}2015}}$
(right), which strongly suggests that -whatever the fit parameters may be- 
total deviations are due to the 
LISW effect at $z<10 $.

\subsection{The LISW in the best fit models.}   
\label{sec:33}

Once that the nature of the deviations between RG and AR-VTG, and how these differences are generated in terms of redshift have been presented, it’s interesting to compare the LISW anisotropies predictions between  best fit models.

So now, instead of use the parameter set values of a LCDM best fit model to be compared with an AR-VTG model that is built from the first one, that is, uses the mentioned parameter set values of the LCDM best fit model, plus a reasonable value for the characteristic AR-VTG parameter (D parameter), the two models to be used are: the LCDM-2013 minimal fit (see first row of Table \ref{tab:2}) and the AR-VTG best fit model presented in Table \ref{tab:1}.

\begin{figure*}[tb]
\begin{center}
\resizebox{1.\textwidth}{!}{%
\includegraphics{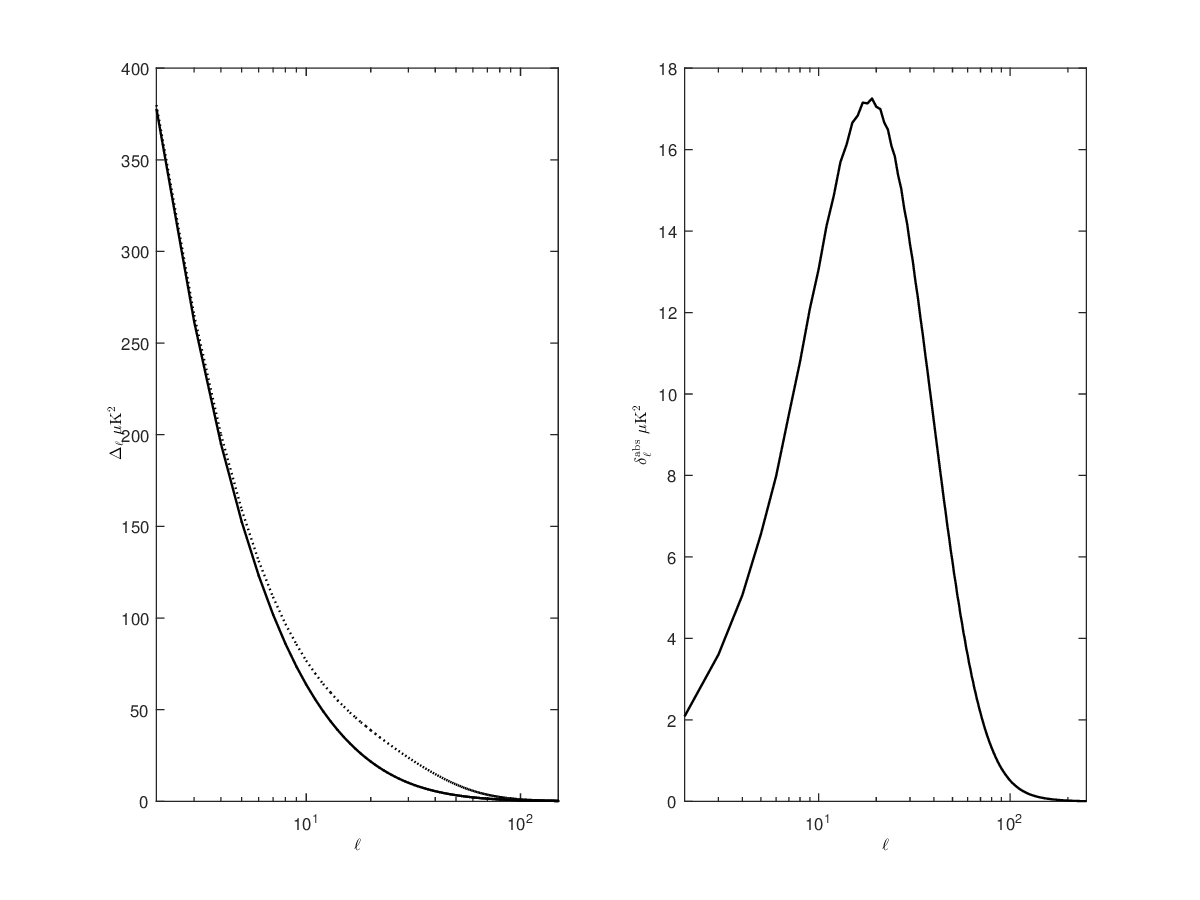}
}
\caption{Presented panels as are those like left and middle in Fig.~\ref{figu2}, but the parameter values correspond to the LCDM 2013 model presented at first row of Table \ref{tab:2} for the solid line, while for the dotted line, the parameters values are those located in Table \ref{tab:1} for the AR-VTG fit.}
\label{figuex03} 
\end{center}      
\end{figure*}

The results of the predictions of the LISW contribution to the CMB anisotropies, of the aforementioned models, are presented in Fig.~\ref{figuex03}. As in previous comparisons, the maximum absolute differences are reached in the range of $\ell$ (10,20), this time those relative deviations are $\sim 50\%$ when examining just the LISW contributions, however, when the total CMB anisotropies are considered, those maximums are between $\sim 1\%$ and $\sim 2\%$. It has been described in section \ref{sec:31} that a set of different results have been obtained when different evidences sources and extended models are studied (see  Refs.~\cite{dal14,dal17} for details), in this global context, we can affirm that the total CMB anisotropies differences reach a maximum between $\sim 2\%$  and  $\sim 5\%$. If we also take into account that, except in the particular case of considering tensor modes and running index, the best fit values obtained for the common parameters of GR and AR-VTG models are very similar, we may conclude that a reasonable doubt exists as to whether CMB has enough discriminating character. Another important issue to take into account is the fact that the low values $\ell$ of the CMB are mainly affected by cosmic variance. In such a case additional cross-correlation tests might be useful. In such a case additional cross-correlation tests might be useful.

The cross-correlation between CMB and some tracers of large scale structure surveys (LSS) was first proposed by Crittenden and Turok (see Ref.~\cite{cri96}), allowing us to isolate the LISW anisotropy contribution. These cross-correlations and certain statistical estimators have been successfully used to provide a physical evidence for dark energy \cite{scr03}, to derive constraints on the dark energy \cite{cor05,sch12} or neutrino masses \cite{les08}, to detect coupling between dark energy and dark matter at low redshifts \cite{oli08}, and other issues. But also provides a mechanism for differentiating dark energy from a modified gravity, even for an identical background expansion \cite{lue04,son07} which is the case we are dealing with.

Although a complete study, based on new cross-correlations, is out of the scope of current paper, we will present below some preliminary results in this regard. With this aim, next we will compare some cross-correlations theoretical predictions for GR and AR-VTG, and for that purpose, let us first introduce and define some suitable useful concepts.

The temperature fluctuation due the ISW effect in a certain direction ${\hat n_1}$ is provided by the expression:

\begin{equation}
\frac{{\Delta T}}{T}\left( {{{\hat n}_1}} \right) =  - 2\int {{e^{ - {\tau _{re}}\left( z \right)}}\frac{{d\Phi }}{{dz}}} \left( {{{\hat n}_1},z} \right)dz ,
\label{eqISW}
\end{equation}

where ${e^{ - {\tau _{re}}(z)}}$ is the visibility function of the photons, and $\Phi$ is the gravitational potential in the Newtonian gauge. The observed density contrast for a certain direction ${\hat n_2}$ is given by:

\begin{equation}
{\delta _g}\left( {{{\hat n}_2}} \right) = \int {{b_g}\left( z \right)\frac{{dN}}{{dz}}} \left( z \right){\delta _m}\left( {{{\hat n}_2},z} \right)dz ,
\label{eqdencon}
\end{equation}

where ${b_g}( z )$ is the galaxy bias, ${{dN}/{dz}}$ is the selection function of the survey, and $\delta_m$ is the matter density fluctuation. For a certain map of CMB anisotropies and a survey of galaxies the cross-correlation and the auto-correlation function are defined as:

\begin{equation}
{C^{Tg}}\left( \theta  \right) \equiv \left\langle {\frac{{\Delta T}}{T}\left( {{{\hat n}_1}} \right){\delta _g}\left( {{{\hat n}_2}} \right)} \right\rangle ,
\label{eqdcorr01}
\end{equation}
and
\begin{equation}
{C^{gg}}\left( \theta  \right) \equiv \left\langle {{\delta _g}\left( {{{\hat n}_1}} \right){\delta _g}\left( {{{\hat n}_2}} \right)} \right\rangle ,
\label{eqdcorr02}
\end{equation}

with the average, denoted by the angular brackets, carried over all the pairs at the equal angular distance  $\theta  = | {{{\hat n}_1} - {{\hat n}_2}}|$. For computing purposes we decompose these quantities using the Legendre polynomials $P_\ell$:

\begin{equation}
{C^{Tg}}\left( \theta  \right) = \sum\limits_{\ell = 2}^\infty  {\frac{{2\ell + 1}}{{4\pi }}C_\ell^{Tg}{P_\ell}} \left[ {\cos \left( \theta  \right)} \right] ,
\label{eqdcorr03}
\end{equation}

the cross-correlation and the autocorrelation power spectra are obtained from:

\begin{equation}
C_\ell^{Tg} = 4\pi \int {\frac{{dk}}{k}{\Delta ^2}} \left( k \right)I_\ell^{ISW}\left( k \right)I_\ell^g\left( k \right) ,
\label{eqdcorr03}
\end{equation}
and
\begin{equation}
C_\ell^{gg} = 4\pi \int {\frac{{dk}}{k}{\Delta ^2}} \left( k \right)I_\ell^g\left( k \right)I_\ell^g\left( k \right) ,
\label{eqdcorr03}
\end{equation}
respectiveliy. The function $\Delta(k)$ is the matter power spectrum, and the two integrals functions $I^{ISW}_\ell$ and $I^{ISW}_\ell$ are defined as follows:

\begin{equation}
I_\ell^{ISW} =  - 2\int {{e^{ - {\tau _{re}}\left( z \right)}}\frac{{d{\Phi _k}}}{{dz}}} {j_\ell}\left[ {kr\left( z \right)} \right]dz ,
\label{eqdcorr04}
\end{equation}
and
\begin{equation}
I_\ell^g = \int {{b_g}\left( z \right)\frac{{dN}}{{dz}}} \left( z \right)\delta _m^k\left( z \right){j_\ell}\left[ {kr\left( z \right)} \right]dz ,
\label{eqdcorr04}
\end{equation}

where $\Phi_k$, $\delta_m^k$ are the Fourier components of the gravitational potential and matter perturbations, respectively, $j_\ell[kr(z)]$ are the spherical Bessel functions and $r(z)$ is the commoving distance at redshift $z$. In order to compute those related quantities a new version of the CROSS-CMBFAST \cite{cor05}, say VTCROSS-CMBFAST, which in turn is an adaptation of the well-known CMBFAST \cite{sel96} code.
In order go ahead with the calculations, some functions are still needed, these are: the galaxy bias $b_g(z)$, and the selection function of the survey $dN/dz$. Let us consider a very simple model with $b_g(z) = 1$ and $dN/dz \sim {z^2}\exp [ { - {{( {10z})}^{1.5}}}]$, that is we select a Gaussian distribution for selection function of the survey as in Ref.~\cite{cor05}. With this set of options and VTCROSS-CMBFAST, the correlation function $C^{Tg}(\theta)$ is calculated for the models: AR-VTG best fit (see Table \ref{tab:1}), $\Lambda$CDM-2013 (row 1 in Table \ref{tab:2}) and AR-VTG with $D = 4 \times 10^8$ (the model introduced in section \ref{sec:32}).

\begin{figure*}[tb]
\begin{center}
\resizebox{1.\textwidth}{!}{%
\includegraphics{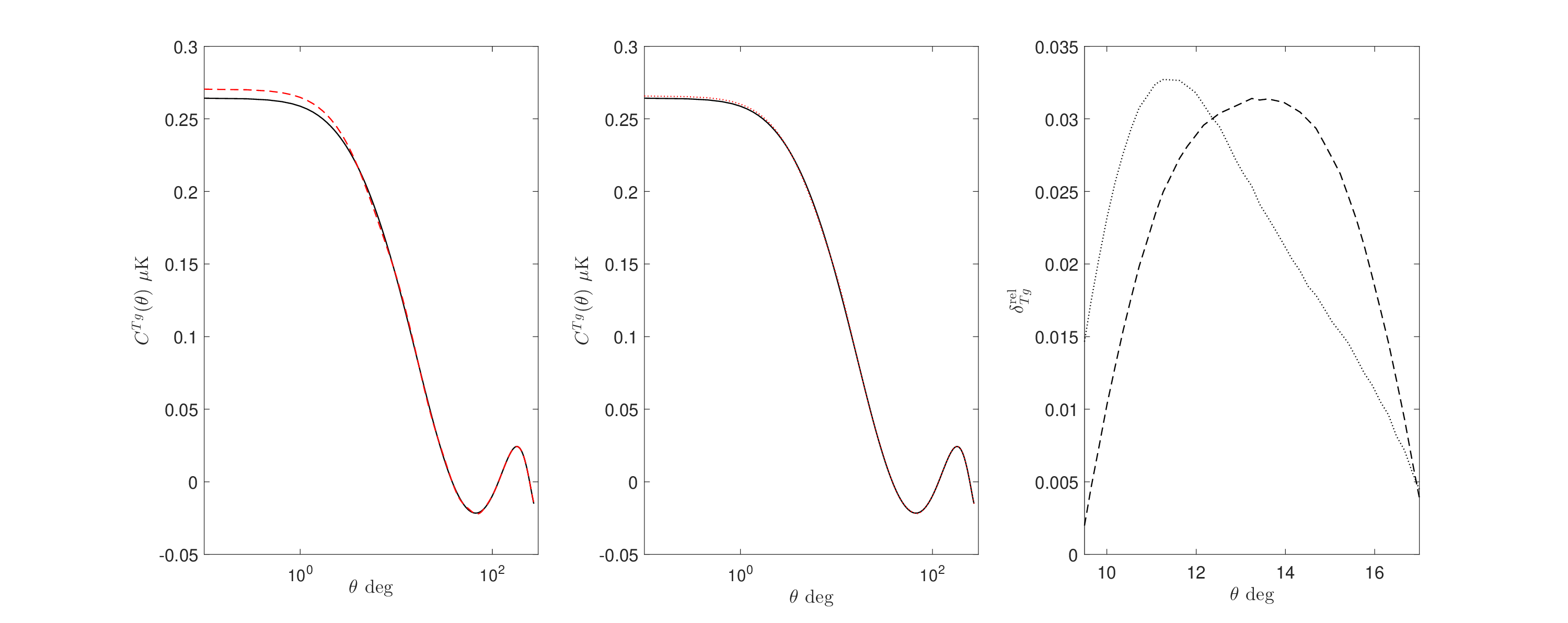}
}
\caption{Left (middle) panel represents the $C_{Tg}(\theta)$ correlations, the solid line has been built using the $\Lambda$CDM-2013 parameters model while the red dashed (dotted) one corresponds to AR-VTG best fit (AR-VTG  with $D = 4 \times 10^8$) settings. In the right panel, dotted (dashed) line provides the relative deviations in the $9 < \theta < 17$ range}
\label{figuex04} 
\end{center}      
\end{figure*}

The results are represented in Fig.~\ref{figuex04}. At this figure, one observes that both models are quite similar; there are relative differences of a $\sim 3\%$ located in the range $9 \lesssim \theta \lesssim 17$, in fact the main contribution to a $C_\ell^{TT}$ multipole corresponds to $\theta = \pi / \ell$, so it's something that might be expected. This is the corresponding $\ell$ range were we found (in the previous section) $C_\ell^{TT}$ deviations between the compared models. The relative differences $\delta_{Tg}^{rel}$ are defined in the same way we did in section \ref{sec:32} for $\delta^{rel}_\ell$.

\section{On the generation of absolute deviations $\delta^{abs}_{\ell}$.}   
\label{sec:4}

\begin{figure*}[tb]
\begin{center}
\resizebox{1.\textwidth}{!}{%
\includegraphics{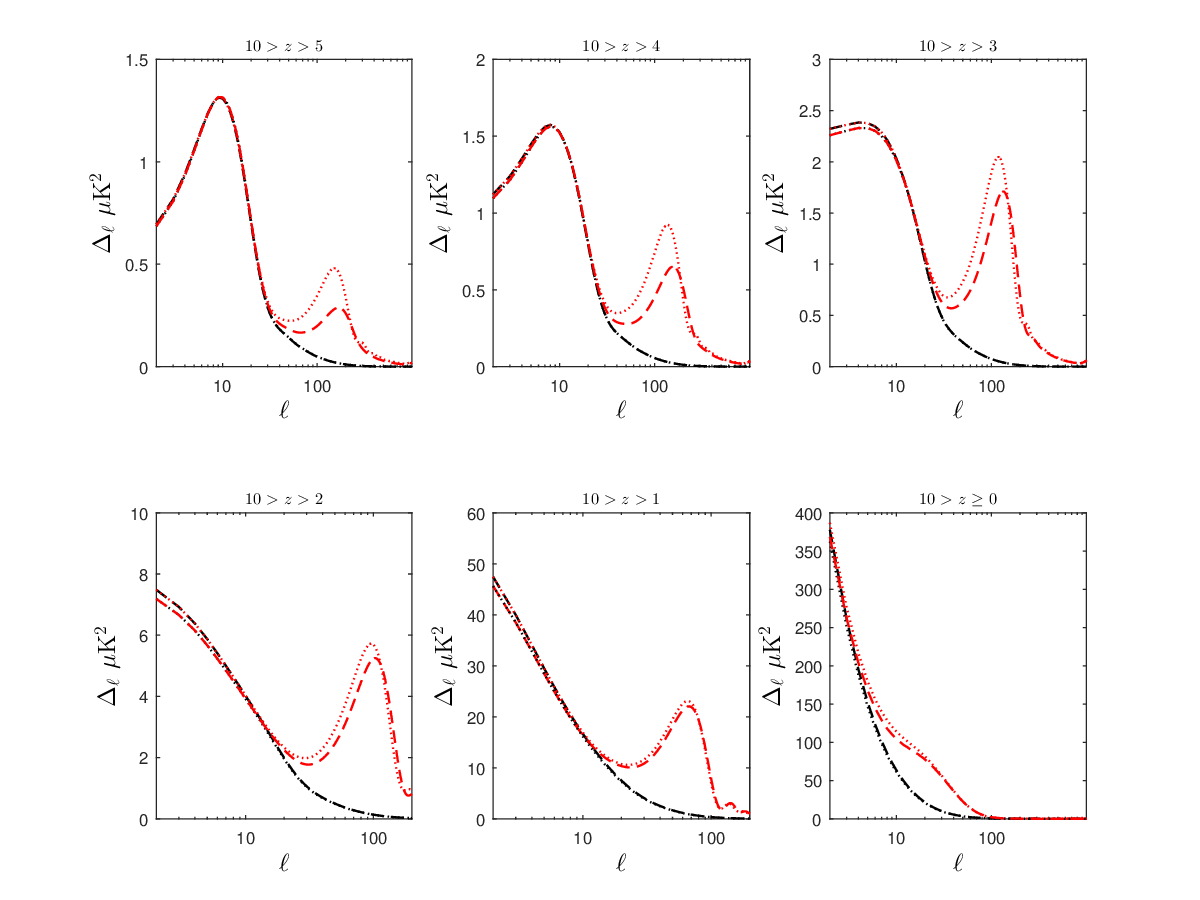}
}
\caption{LISW contributions to $\Delta_{\ell}$ generated 
in the redshift intervals indicated above the panels. There are two 
black and two red curves in each panel. Black (red) dashed lines show
LISW contributions to $\Delta_{\ell}$ for $\Lambda {\rm{CDM{-}2013}}$ in
the $\Lambda$CDM model of GR (AR-VTG with $D=4 \times 10^{8}$).
Same for black and red dotted lines in the $\Lambda {\rm{CDM{-}2015}}$
parameter configuration.
}
\label{figu4} 
\end{center}      
\end{figure*}

After analysing the nature of the differences between the CMB anisotropies 
in GR and AR-VTG, let us study how these differences are generated between 
redshifts 10 and 0. To do that, the differences have been estimated while the redshift varies from 10, to 5, 4, 3, 2, 1 and 0. 
Results are presented in Fig.~\ref{figu4}, where each panel corresponds to one of the above 
redshift variations, which is given above the panel.

For the $\Lambda {\rm{CDM{-}2013}}$ parameters, the black dashed lines ($\Lambda$CDM model of GR)
must be compared with the red dashed lines (AR-VTG with $D=4 \times 10^{8}$). The 
separation between these lines measures the differences between the CMB anisotropies 
in GR and AR-VTG. Top left panel shows that, for $10 > z > 5$, these $\Delta_{\ell} $ 
differences reach only a few tenths of $\mu K^{2} $, and for $10 > z > 3$ (top right panel) 
a few $\mu K^{2} $; hence, we can conclude that the differences are essentially 
generated for $3 > z > 0$. See bottom panels for details.

The same may be concluded, for the $\Lambda {\rm{CDM{-}2015}}$ parameters,
from the comparison between black dotted lines ($\Lambda$CDM model of GR) and 
red dotted lines (AR-VTG with $D=4 \times 10^{8}$); hence, the conclusion that 
the differences between the CMB anisotropies 
in GR and AR-VTG is mainly generated for $3 > z > 0$, is a robust conclusion 
almost independent on the selected parameters.


\section{Discussion and conclusions.}   
\label{sec:5}

Our main conclusions are the following:
The total absolute $\Delta_{\ell}$ deviations, between 
the $\Lambda$CDM model of GR and AR-VTG with $D=4 \times 10^{8}$,
are due to the LISW effect (including $R1$). These deviations are essentially produced 
between redshifts 3 and 0 with the main part generated for $z \leq 1$. 
The relative deviations are close to 6\% for $10 \leq \ell \leq 20$ and 
greater than 1\% for $2 \leq \ell \leq 60$.

\begin{figure*}[tb]
\begin{center}
\resizebox{1.\textwidth}{!}{%
\includegraphics{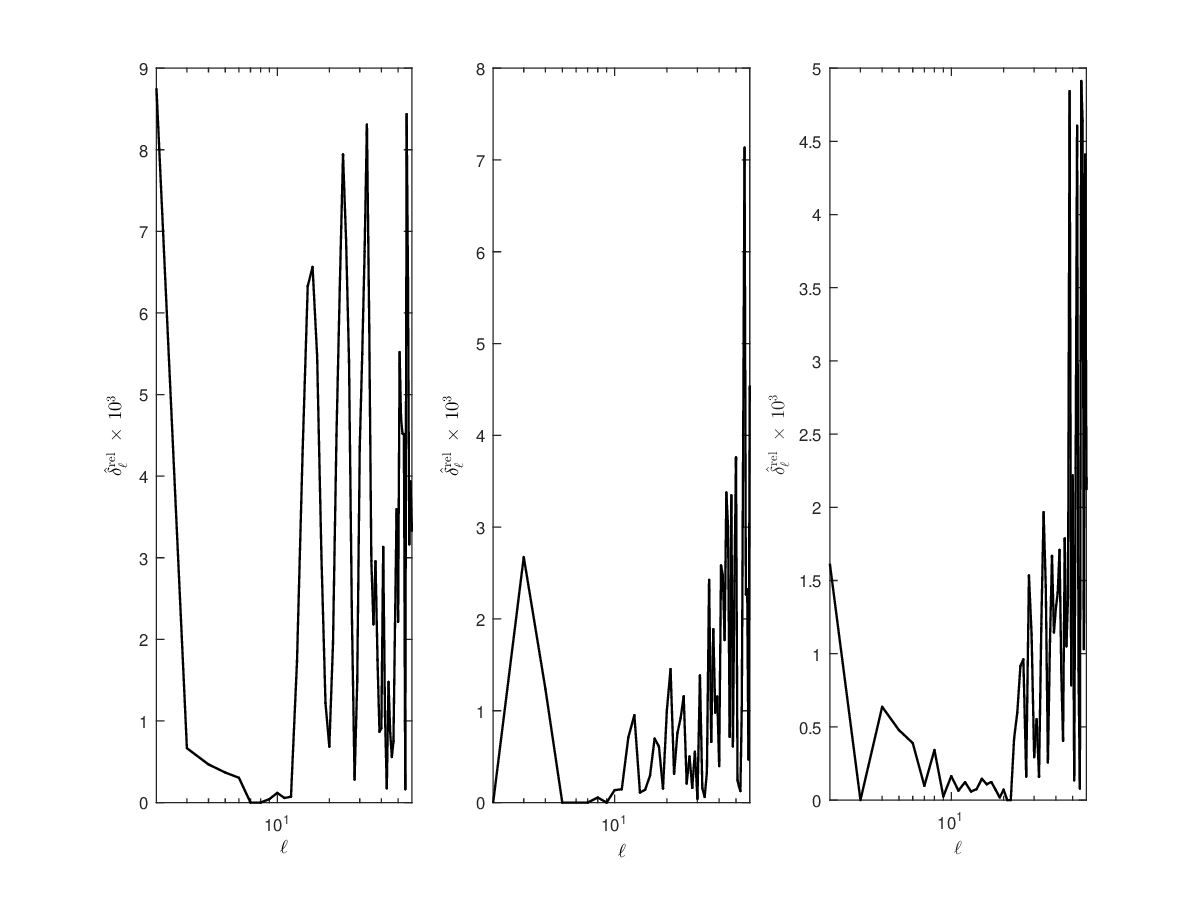}
}
\caption{Representation of the $\hat{\delta}^{rel}_{\ell}$ quantities in terms of $\ell$. 
Left: for the left bottom panel of Fig.~\ref{figu1}. Centre: for the
central panel of Fig.~\ref{figu2}. Right: for the right bottom panel of Fig.~\ref{figu1}.
}
\label{figu5} 
\end{center}      
\end{figure*}

Up to now, the nature of the aforementioned deviations has been suggested by the fact that
the dotted and dashed lines of three panels are almost identical to the eye. These three 
panels are the left bottom panel of Fig.~\ref{figu1} ($\Lambda {\rm{CDM{-}2013}}$ parameters with 
reionization), the central panel of Fig.~\ref{figu2} ($\Lambda {\rm{CDM{-}2013}}$ parameters without 
reionization), and the right bottom panel of Fig.~\ref{figu1} ($\Lambda {\rm{CDM{-}2015}}$ parameters with 
reionization). Let us now prove that the relative differences, $\hat{\delta}^{rel}_{\ell}$, between the dotted and the 
dashed lines of each of these three panels are smaller than the relative errors of CAMB and VTCAM
calculations, which, based on our hard numerically computational tests and the settings we fix in terms of a balance between
accuracy and performance, may be estimated to be around 1\%. These relative differences are 
$\hat{\delta}^{rel}_{\ell} = 2 [|\delta^{abs}_{\ell} (dot)- \delta^{abs}_{\ell} (dash)|]/
[\delta^{abs}_{\ell} (dot)+ \delta^{abs}_{\ell} (dash)] $. Quantities $\hat{\delta}^{rel}_{\ell}$ 
are given in the three panels of Fig.~\ref{figu5}. In any case, these quantities are 
smaller than $0.001$ and, consequently, they are smaller than the numerical errors.

The LISW effect, relevant for $\ell \leq 100 $, is produced at low redshifts and, consequently, it may be detected by looking for cross correlations 
between the CMB temperature distribution and tracers of the dark matter distribution on large spatial scales \cite{cri96}.
This kind of detection has been recently achieved by using Planck data and appropriate tracers; see Ref.~\cite{planck2014_19},
where it is claimed that some detected cross-correlations are compatible with the $\Lambda$CDM predictions; although other models may also be admissible. 
Previous detections are also listed in Ref.~\cite{planck2014_19}.

Since the LISW effects are distinct in the $\Lambda$CDM model of GR and in AR-VTG with $D=4 \times 10^{8} $,
with small relative differences (reaching values close to 6\%), as it has been mentioned at the beginning of this section,
the cross correlations predicted in the contexts of both models should be also different
although comparable and; then, the question is: Could we select one of these models by comparing the 
cross-correlations predicted in them with those observed? 
Could we do that with high statistical significance? In section \ref{sec:31} we have outlined what could be a continuity way for this research, 
in such a case, we should use more complex models and estimators. A deep study on this line is out of the scope of this paper, 
but could lead to the selection of one of the AR-VTG models in future.

\section*{Acknowledgments}

This work has been supported by the Spanish
Ministry of {\em Econom\'{\i}a y Competitividad},
MINECO-FEDER project FIS2015-64552-P and CONSOLIDER-INGENIO project 
CSD2010-00064. Calculations were carried out at the Centre de C\`alcul de la Universitat de Val\`encia.
On behalf of myself and my colleague Diego (R.I.P.), 
I want to thank Pier-Stefano Corasaniti which has generously provide the CROSS-CMBFAST original code.


\end{document}